\documentclass[prb,twocolumn,showpacs,preprintnumbers,amsmath,amssymb,superscriptaddress,aps]{revtex4-1}

\usepackage{amsmath,amssymb,amsfonts} 	
\usepackage{graphicx}
\usepackage{hhline}
\usepackage[latin1]{inputenc}
\usepackage{subfigure}

\begin{document}

\title{Assessing the performance of the Random Phase Approximation for exchange and superexchange coupling constants in magnetic crystalline solids}

\author{Thomas Olsen}
\email{tolsen@fysik.dtu.dk}
\affiliation{Computational Atomic-Scale Materials Design (CAMD) and Center for Nanostructured Graphene (CNG), Department of Physics, Technical University of Denmark}

\begin{abstract}
The Random Phase Approximation (RPA) for total energies has previously been shown to provide a qualitatively correct description of static correlation in molecular systems, where density functional theory (DFT) with local functionals are bound to fail. This immediately poses the question of whether the RPA is also able to capture the correct physics of strongly correlated solids such as Mott insulators. Due to strong electron localization, magnetic interactions in such systems are dominated by superexchange, which in the simplest picture can be regarded as the analogue of static correlation for molecules. In the present work we investigate the performance of the RPA for evaluating both superexchange and direct exchange interactions in the magnetic solids NiO, MnO, Na$_3$Cu$_2$SbO$_6$, Sr$_2$CuO$_3$, Sr$_2$CuTeO$_6$, and a monolayer of CrI$_3$, which are chosen to represent a broad variety of magnetic interactions. It is found that the RPA can accurately correct the large errors introduced by Hartree-Fock - independent of the input orbitals used for the perturbative expansion. However, in most cases, accuracies similar to RPA can be obtained with DFT+U, which is significantly simpler from a computational point of view.
\end{abstract}
\maketitle

\section{Introduction}
The Random Phase Approximation (RPA) was introduced in the early 1950s by Bohm and Pines, as a means to provide a collective description of interacting electrons.\cite{Bohm1951,Pines1952,Bohm1953} The approach naturally incorporates long range correlations and thus provides a qualitative account of collective excitations in metallic systems - the plasmons. A few years later it was shown that the RPA could be understood as a resummation of the most divergent terms emerging in a perturbative treatment of the Coulomb interaction and an analytical expression for the RPA correlation energy of the interacting electron gas was derived by Gell-Mann and Brueckner.\cite{Gellmann} Whereas, the second order perturbative correction to the total energy of the homogeneous electron gas gives rise to a divergence, the full RPA resummation can be interpreted as a renormalization of the Coulomb interaction due to screening of itinerant electrons and provides a finite total energy. Nevertheless, while the significance of RPA for understanding many-body phenomena cannot be overestimated, the accuracy of the approximation for total energies is somewhat unsatisfactory. A comparison with Quantum Monte Carlo simulations,\cite{Ceperley1980} shows that RPA underestimates the total energy per electron by approximately 0.5 eV per electron for the homogeneous electron gas with densities corresponding to those of common metals.\cite{Lein2000, Olsen2012}

The perturbative evaluation of correlation energies within RPA is easily generalized to non-uniform systems. In general, the correlation energy can be written as a functional of the interacting density-density response function,\cite{Langreth1975} which can be calculated from the non-interacting density-density response function within RPA. For the homogeneous electron gas it is natural to use non-interacting orbitals and eigenvalues as input for the evaluation of the non-interacting response function. However, for typical non-uniform systems, the bare non-interacting orbitals usually provides a poor starting point for perturbation theory and much better results can be obtained if either Hartree-Fock or Kohn-Sham orbitals are used. While such a scheme generally provides more reliable results, it also means that the calculated RPA correlation energy becomes dependent on the choice of input orbitals or - in the context of the DFT - the choice of exchange-correlation functional used to generate the input orbitals. Alternatively, it is possible to define a local RPA exchange-correlation potential from the total energy expression by means of the optimized effective potential approach\cite{Godby1988, Kotani1999} and thus obtain the selfconsistent RPA total energy using the framework of DFT. However, for most applications the selfconsistent approach is prohibitively demanding in terms of computational power and one must resort to the perturbative evaluation of RPA correlation energies, which then acquire a dependence on input orbitals.

In general, the absolute RPA correlation energy is always severely underestimated - typically on the order of 0.5 eV per electron. Nevertheless, due to the universal nature of this self-correlation error, there is a large degree of error cancellation whenever one is calculating total energy {\it differences}. The non-selfconsistent RPA scheme has been applied to molecular atomization energies,\cite{Furche2001,Ren2012,Olsen2013,Olsen2014a} barrier heights,\cite{Paier2012,Ren2012} cohesive energies of solids,\cite{Harl2010a,Kresse2009,Yan2013,Olsen2013,Olsen2014a,Patrick2016} adsorption energies,\cite{Ren2009,Schimka2010,Olsen2013,Olsen2014a,Jauho2015} van der Waals bonded systems,\cite{Marini2006,Harl2008,Lebegue2010,Olsen2011,Mittendorfer2011,Ren2012,Bjorkman2012,Olsen2013,MacHer2014} and dissociation of small molecules.\cite{Furche2001,Ren2009,Henderson2010,Hellgren2012,Mori-Sanchez2012,Olsen2013,Olsen2014} To summarize the general trends, RPA tends to perform slightly worse than PBE for covalent bonds. For example, RPA gives a mean absolute relative error of 7 {\%} for the cohesive energies of solids whereas PBE gives a 5 {\%} error and RPA and PBE both gives an error of 6 {\%} for the atomization energies of molecules - typically corresponding to a total error of 0.5-1.0 eV for small molecules.\cite{Olsen2014a} In contrast, the non-local nature of the RPA total energy functional naturally encompasses dispersive interactions and RPA thus provides a good account of van der Waals interactions although small systematic errors have been reported.\cite{Ren2011} A rather surprising property of the RPA for the energies, is the qualitatively correct description of molecular dissociation of closed shell molecules. In the case of N$_2$\cite{Furche2001} and H$_2$ \cite{Ren2012, Eshuis2012, Olsen2013} RPA accurately reproduces the strict dissociation limit if one compensates for the self-correlation error of the individual atoms. This limit comprises a prime example of static correlation, where the input Kohn-Sham Slater determinant is a very poor approximation to the true ground state wavefunction and the dissociation cannot be correctly described by common semi-local functionals such as LDA and PBE.

The fact the RPA can account for static correlation in simple molecules provides hope that RPA may give a good description of the ground state energy of strongly correlated materials in general. In particular, static correlation is known to play an important role is Mott insulators such as NiO and MnO and Kohn-Sham DFT usually predicts such materials to be metals or small gap semiconductors. This is not a problem in itself since DFT can only be expected to reproduce the ground state density and energy. However, it is highly challenging for any approximate exchange-correlation functional to reproduce the correct ground state properties based on a Kohn-Sham Slater determinant that do not resemble the true many-body wavefunction. A common feature in many of these materials is an antiferromagnetic ground state resulting from superexchange interactions between spin states localized on the transition metal atoms.\cite{Anderson1950a, Anderson1959} In general the magnetic structure can be described in terms of Heisenberg Hamiltonians, which are characterized by a set of exchange couplings $J_{ij}$. These coupling constants represent the exchange interactions between spins localized at atoms $i$ and $j$ and knowledge of the $J_{ij}$ will allow one to calculate various observables such as the magnetic susceptibility, transition temperature and magnetic excitation spectra. In principle, the $J_{ij}$ can be calculated by comparing ground state energies of different spin configurations\cite{Rosner1997, Illas1998} and should thus be accessible by standard DFT calculations. In practice, however, it is often challenging for standard semi-local functionals due to the static correlation inherent in many magnetic materials.

In this paper we investigate the performance of the RPA for evaluating the magnetic coupling constants $J_{ij}$ in magnetic materials. We compare with LDA + U,\cite{Liechtenstein1995} PBE + U, and HSE06\cite{Krukau2006} calculations and assess the performance for the materials NiO, MnO, Sr$_2$CuO$_3$, SrCuTeO$_6$, Na$_3$Cu$_2$SbO$_6$, and a monolayer of CrI$_3$, which are chosen in order to represent a broad class of magnetic interactions. We evaluate the magnetic coupling constants for various values of the on-site Coulomb repulsion U with either LDA, PBE, or HSE06, and show that it is often possible to obtain good agreement with experiment if U is chosen "correctly". In contrast RPA, is rather insensitive to the choice of $U$ used to obtain the input orbitals and provides good agreement with experiments for a wide range of U-values.  It should be noted that the term RPA is also used in spin-wave theory, where it comprises a simple means to include interactions between magnons.\cite{Yosida1996} In the present work, however, RPA will exclusively denote the perturbative treatment of total electronic energies.

The paper is organized as follows. In Sec. \ref{sec:theory} we lay out the foundations necessary to evaluate RPA total energies and summarize how magnetic coupling constants can be obtained from the energy mapping method. In Sec. \ref{sec:results} we present the results of the computations and in Sec. \ref{sec:discussion} we provide a brief discussion and outlook.

\section{Theory}\label{sec:theory}
\subsection{Random phase approximation}
The RPA for total energies is straightforward to derive from the adiabatic connection and fluctuation-dissipation theorem. Briefly, the correlation energy within DFT can be written as
\begin{equation}\label{eq:E_c}
 E_c=-\int_0^1d\lambda\int^\infty_0\frac{d\omega}{2\pi}\mathrm{Tr}[v_c\chi^\lambda(i\omega)-v_c\chi^{KS}(i\omega)],
\end{equation}
where $v_c$ is the Coulomb interaction and $\chi^\lambda(i\omega)$ is the response function of an interacting system where the Coulomb interaction has been rescaled by $\lambda$ evaluated, at the imaginary frequency $i\omega$. In the context of time-dependent density functional theory, the RPA response function can be derived from the Dyson equation
\begin{equation}\label{eq:dyson}
\chi^\lambda_{\mathrm{RPA}}(i\omega)=\chi^{KS}(i\omega)+\chi^{KS}(i\omega)\lambda v_c\chi^\lambda_{\mathrm{RPA}}(i\omega).
\end{equation}
Inserting the solution of Eq. \eqref{eq:dyson} into into Eq. \eqref{eq:E_c} and carrying out the $\lambda$ integration then yields
\begin{equation}\label{eq:E_c^RPA}
 E_c^{\mathrm{RPA}}=\int^\infty_0\frac{d\omega}{2\pi}\mathrm{Tr}\Big\{\ln[1-v_c\chi^{KS}(i\omega)]+v_c\chi^{KS}(i\omega)\Big\}.
\end{equation}

For solid state systems it is most convenient to expand the wave-functions in a basis of plane waves $\sim e^{i\mathbf{G}\cdot\mathbf{r}}$, where $\mathbf{G}$ is a reciprocal lattice vector. In this basis the non-interacting response function can be written as
\begin{align}\label{eq:chi^KS}
\chi^{KS}_{\mathbf{G}\mathbf{G}'}&(\mathbf{q},i\omega)=\frac{1}{V}\sum_{m,n}\sum_{\mathbf{k}\in\mathrm{BZ}}\frac{f_{n\mathbf{k}}-f_{m\mathbf{k+q}}}{i\omega+\varepsilon_{n\mathbf{k}}-\varepsilon_{m\mathbf{k+q}}}\\
&\times\langle\psi_{n\mathbf{k}}|e^{-i(\mathbf{q+G})\cdot\mathbf{r}}|\psi_{m\mathbf{k+q}}\rangle\langle\psi_{m\mathbf{k+q}}|e^{i(\mathbf{q+G'})\cdot\mathbf{r}}|\psi_{n\mathbf{k}}\rangle\notag,
\end{align}
where $f_{n\mathbf{k}}$ and $\varepsilon_{n\mathbf{k}}$ are respectively the occupation factor and eigenenergy of the Bloch state $|\psi_{n\mathbf{k}}\rangle$. In this representation the trace in Eq. \eqref{eq:E_c^RPA} becomes a Brillouin zone integral over $\mathbf{q}$ in addition to the trace of $\mathbf{G}$-vectors. 

Compared to standard semi-local functionals the RPA correlation energy is much more computationally demanding. The two main reasons for this are: 1) The RPA correlation energy is a functional of the two-point function $\chi^{KS}_{\mathbf{G}\mathbf{G}'}$ rather than the single-point density $n_\mathbf{G}$, which is sufficient for semi-local functionals. 2) The non-interacting response function depends  on the unoccupied bands as well as the occupied ones whereas explicit density functionals only depend on the occupied orbitals. For large systems, absolute convergence with respect to plane waves and unoccupied states becomes unfeasible and one has to resort to an extrapolation scheme in order to obtain converged results. It has previously been demonstrated that the energy as a function of cutoff scales as\cite{Harl2008, Shepherd2012, Olsen2013}
\begin{align}\label{eq:E_cut}
E^{\mathrm{RPA}}(E_{\mathrm{cut}})=E^{\mathrm{RPA}}+\frac{A}{E_{\mathrm{cut}}^{3/2}},
\end{align}
where $E_{\mathrm{cut}}$ is an energy cutoff determining how many plane waves are included and the number of states included in the summation entering Eq. \eqref{eq:chi^KS}. With this expression the converged energy $E^{\mathrm{RPA}}$ can be computed accurately by extrapolation.

\subsection{Heisenberg model}
The Heisenberg Hamiltonian can be derived as a low energy approximation to the full many-body Hamiltonian using first order perturbation theory in the Coulomb interaction as a starting point. Except for a spin-independent constant the Hamiltonian then becomes
\begin{align}\label{eq:Heisenberg}
H=-\frac{1}{2}\sum_{ij}J_{ij}\mathbf{S}_i\cdot\mathbf{S}_j,
\end{align}
where $\mathbf{S}_j$ is the total spin operator for site $i$ and 
\begin{align}\label{eq:exchange}
J_{ij}^\mathrm{exc}=\frac{2}{S^2}\sum_{n_in_j}\int&\frac{d\mathbf{r}d\mathbf{r}'}{|\mathbf{r}'-\mathbf{r}|}\varphi^*_{n_i}(\mathbf{r})\varphi^*_{n_j}(\mathbf{r}')\varphi_{n_i}(\mathbf{r}')\varphi_{n_j}(\mathbf{r})
\end{align}
is the exchange integral with $S$ being the maximal allowed eigenvalue of $\mathbf{S}$. In Eq. \eqref{eq:exchange} it was assumed that the magnetic moment at lattice site $i$ is comprised of the localized orbitals $\varphi_n$ and the sum runs over the occupied orbitals if the $\varphi_n$ shell is less than half filled and over unoccupied orbitals if the $\varphi_n$ shell is more than half filled.\cite{Yosida1996} It is readily verified that the exchange integral \eqref{eq:exchange} is strictly positive.

However, the model \eqref{eq:Heisenberg} may also be derived from a completely different point of view. In materials with strongly localized orbitals it will be a better approximation to start with the atomic limit of lattice sites with addition and removal energies $U$ on the individual sites. Including inter-site hybridization perturbatively then yields the model \eqref{eq:Heisenberg}, but with the coupling parameters given by
\begin{align}\label{eq:se}
J_{ij}^\mathrm{se}=-\frac{4}{S^2}\sum_{n_in_j}\frac{|t_{n_in_j}|^2}{U},
\end{align}
where $t_{n_in_j}$ is the hopping integral between orbitals $\varphi_{n_i}(\mathbf{r}-\mathbf{R}_i)$ and $\varphi_{n_j}(\mathbf{r}-\mathbf{R}_j)$. The physical origin of $J_{ij}^\mathrm{exc}$ and $J_{ij}^\mathrm{se}$ are very different; whereas the exchange coupling $J_{ij}^\mathrm{exch}$ originates from a first order perturbative treatment of the Coulomb interaction, $J_{ij}^\mathrm{se}$ is strictly non-perturbative in the on-site Coulomb interaction $U$. The coupling constants $J_{ij}^\mathrm{se}$ are typically mediated by non-magnetic anions and the mechanism is denoted superexchange.\cite{Anderson1950a,Anderson1959}

In general, both exchange and superexchange may contribute to the $J_{ij}$ and we may consider the model \eqref{eq:Heisenberg} with
\begin{align}\label{eq:J}
J_{ij}=J_{ij}^\mathrm{exc}+J_{ij}^\mathrm{se}.
\end{align}
Finally, we note that the Hamiltonian \eqref{eq:Heisenberg} may contain other terms describing magnetic anisotropy, Dzyaloshinskii-Moriya interaction,\cite{Dzyaloshinsky1958, Moriya1960a} biquadratic exchange and four-spin interactions.\cite{Yosida1996} However, these terms are usually less significant and in the present work we will only be concerned with the determination of the $J_{ij}$ from first principles.

\subsection{Energy mapping scheme}
In principle one could try to obtain the coupling parameters in the model Hamiltonian \eqref{eq:Heisenberg} from Eqs. \eqref{eq:exchange} and \eqref{eq:se}. Starting with a set of Kohn-Sham Bloch states, one would be required to construct Wannier functions and then evaluate the relevant Coulomb and hopping integrals. Even if this procedure could be completed, it may not be very accurate due to the explicit dependence on Kohn-Sham orbitals, which are not required to give a faithful representation of the true many-body wavefunction. Instead, the ground state energy for a given spin configuration should be accurately described within DFT - at least if a good approximation to the exchange-correlation functional is applied.

We thus choose a spin configuration $l$ and fix the spin state of the lattice sites $S^l_i$ accordingly. Since, we are only considering collinear spin configurations the $S_i$ can take the values of $\pm S$. We thus obtain a set of equations
\begin{align}\label{eq:E_map}
E^l=-\frac{1}{2}\sum_{ij}J_{ij}S^l_iS_j^l,
\end{align}
which are solved for the $J_{ij}$.\cite{Xiang2013} In order to determine $N$ parameters $J_{ij}$, we need $N+1$ {\it ab initio} calculations to obtain the needed $E^l$. The spins in the DFT calculations are unconstrained in the sense that only an initial spin configuration is provided and the spins and density is thus relaxed until a local minimum of the total energy is reached. Usually the local minimum will retain the qualitative spin configuration that was provided and one obtains $E^l$ for a well-defined spin configuration. However, it may happen that the spin configuration is lost, such that the minimization does not find a local minimum and one has to try another configuration.

When solving Eqs. \eqref{eq:E_map} for the parameters, we always use an exact half-integer for the total spin $S$ based on the oxidation number of the magnetic ion. It is possible to extract a measure of the effective spin residing on a magnetic ion based on the DFT calculations, but the construction is somewhat arbitrary and we have thus chosen to use the full spin here for consistency. More precisely, the oxidation number of an atom in an insulting solid can be defined rigorously based on the Berry phase.\cite{Jiang2012b} and this uniquely determines the spin of an atom using Hunds rule. However, the Berry phase calculation is rather involved and here we have simply chosen to extract the spin of the magnetic ions based on the ferromagnetic electronic structure of the Kohn-Sham system, which always yields an integer number of Bohr magnetons for an insulator.

\subsection{Calculational details}
All calculations was performed with the electronic structure software package GPAW,\cite{Enkovaara2010a} which is based on the projector augmented wave method.\cite{blochl} combined with the Atomic Simulation Environment (ASE).\cite{Larsen2017} The LDA+U and PBE+U calculations are self-consistent whereas the HSE06 and RPA calculations are performed on top of PBE+U. The calculations was performed in plane wave mode and all densities and wavefunction was generated from spin-polarized collinear input calculations with a cutoff of 600 eV. For the RPA calculations we applied a two-point extrapolation using Eq. \eqref{eq:E_cut} with $E_\mathrm{cut}=114$ eV and $E_\mathrm{cut}=150$ eV. With these cutoff energies, the energy {\it differences} between various spin states are accurately converged and for all systems we tested that a two-point extrapolation from a different pair of cutoff energies yielded the same energy difference to less than one percent accuracy. The k-point meshes applied to the individual materials will be stated below. We note that the exact exchange calculations for some materials required rather tight convergence criteria on the Kohn-Sham orbitals and we used a criteria of $10^{14}$ eV$^2$ per valence electron for the integrated value of the square of the residuals of the Kohn-Sham equations in order to obtain convergence.

\section{Results}\label{sec:results}

\subsection{NiO and MnO}
The transition metal oxides comprises prototypical Mott insulators with a charge transfer gap and antiferromagnetic ground states.\cite{Cox2010} The oxygen ions are $O^{-2}$ and for NiO and MnO the transition metal atoms thus constitute $S=1$ and $S=5/2$ systems originating from the $3d^8$ and $3d^5$ shells respectively. In both systems the ground state is comprised of ferromagnetic planes that are antiferromagnetically coupled. We will use this case to exemplify the energy mapping scheme. Each transition metal atom has 12 equivalent nearest neighbors and 6 equivalent next nearest neighbors. A given transition metal ion in the antiferromagnetic ground state has all next nearest atoms anti-aligned, six nearest neighbors aligned, and six nearest neighbors anti-aligned. The nearest neighbor magnetic coupling is denoted by $J_1$ and the next nearest neighbor coupling - mediated by an intermediate oxygen atom - by $J_2$. In order to compute these parameters we consider three spin configurations: the antiferromagnetic (AF) ground state, a completely ferromagnetic state (F) and another antiferromagnetic state (AF') that differs from the ground state in that all spins connected by an oxygen bridge are aligned and only four of the nearest neighbors are aligned. Although the antiferromagnetic ground state can be obtained with a unit cell containing two Mn/Ni atoms, the AF' configuration cannot be described in this cell and we used a cubic cell with 8 Mn/Ni atoms for all calculations in order to minimize errors in the energy differences and a gamma-centered k-point grid of $4\times4\times4$.

For a single Mn/Ni atom the three energies can then be written
\begin{align}
E^{AF}&=-\frac{1}{2}\Big(-6S^2J_2\Big),\label{eq:E_map1}\\
E^{F}&=-\frac{1}{2}\Big(12S^2J_1+6S^2J_2\Big),\label{eq:E_map2}\\
E^{AF'}&=-\frac{1}{2}\Big(-4S^2J_1+6S^2J_2\Big)\label{eq:E_map3},
\end{align}
and the couplings are thus
\begin{align}\label{eq:J_map}
J_1&=\frac{1}{8S^2}\Big(E^{AF'}-E^{F}\Big),\\
J_2&=-J_1-\frac{1}{6S^2}\Big(E^{F}-E^{AF}\Big).
\end{align}

\begin{figure}[t]
   \includegraphics[width=4.2 cm]{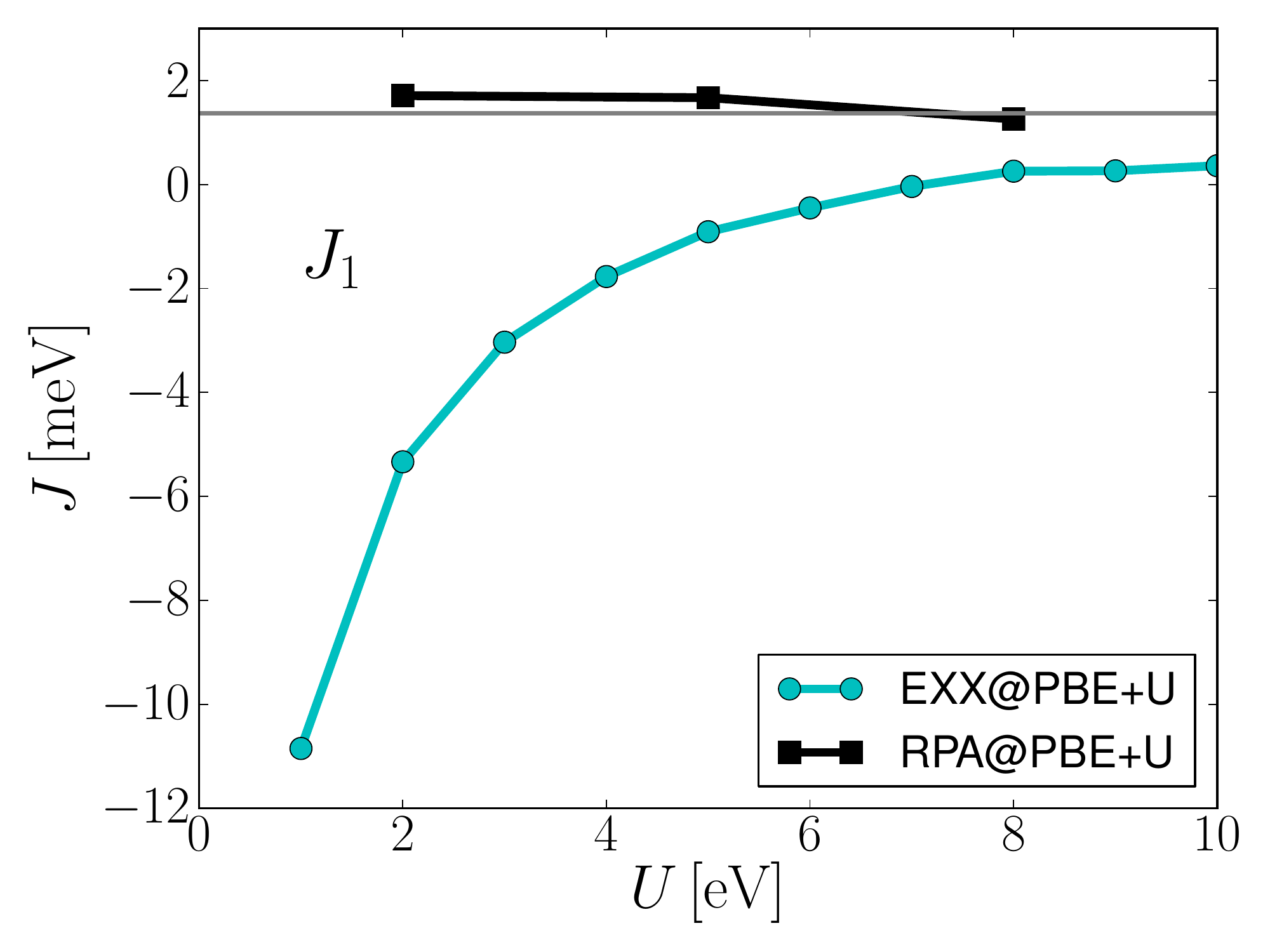}
   \includegraphics[width=4.2 cm]{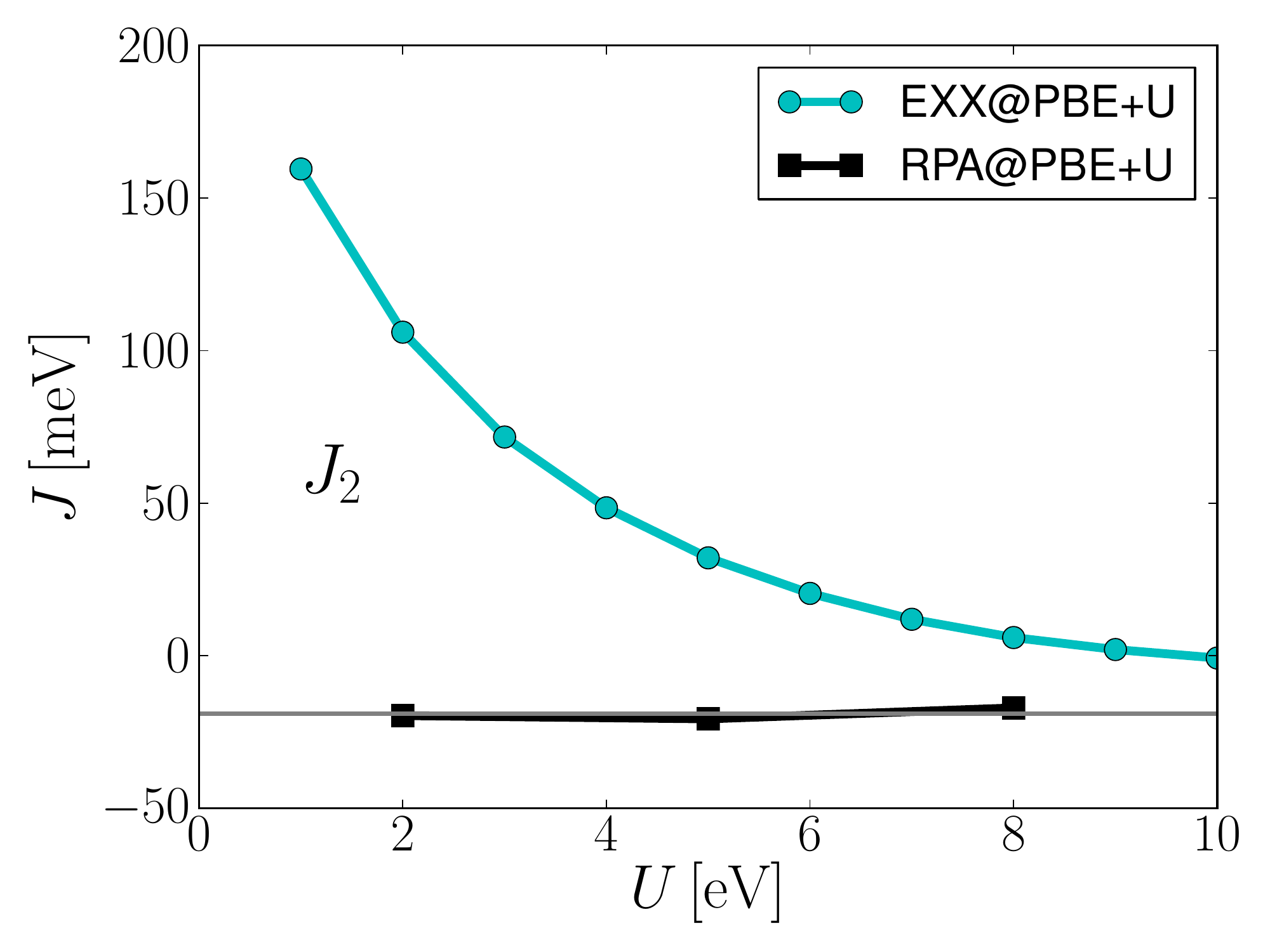}
\caption{(Color online) $J_1$ and $J_2$ parameters for NiO calculated with EXX and RPA calculated as a function of U used in the input PBE+U calculations. The experimentally determined values are shown as a horizontal grey line.}
\label{fig:EXX}
\end{figure}

In Fig. \ref{fig:EXX} we display exact exchange (EXX) and RPA calculations for NiO performed non-selfconsistently on top of PBE +U orbitals and eigenvalues. The coupling constants $J_1$ and $J_2$ has been determined experimentally from inelastic neutron scattering measurements\cite{Hutchings1972} and is shown as grey horizontal lines. It is evident that EXX is strongly dependent on the input orbitals - for small values of U, the orbitals are rather delocalized and EXX is very far from the experimental results. This is indeed expected since the the ground state of NiO is dominated by superexchange effects that anti-align all Ni atoms connected by an oxygen bridge and this is reflected in the fact that $J_2$ is an order of magnitude larger than $J_1$. Since superexchange is far more important in this case, we expect that the correlation energy must play a significant a role and we observe that RPA is able to accurately correct the errors introduced by EXX for all the applied values of U. We note that due to the linearity of the energy mapping scheme we can write
\begin{align}\label{eq:J_rpa}
J_{ij}=J_{ij}^{EXX}+J_{ij}^{RPA},
\end{align}
where $J_{ij}^{RPA}$ is a correction coming from the correlation energy alone. From Fig. \ref{fig:EXX} it is then clear that RPA provides a large correction for small values of U and a small correction for large values of U. It is highly remarkable that a perturbative treatment within RPA is almost able to completely eliminate the initial state dependence. A similar result was noted in Ref. \onlinecite{Patrick2016}, where the RPA energy difference between anatase and rutile TiO$_2$ was shown to be nearly independent of the value of U used to generate input orbitals. Unfortunately it was not possible to extend the calculations to U=0. Although the PBE ground state has a (small) gap, the ferromagnetic state used for the energy mapping becomes metallic for U $<$ 1.0 eV and it becomes highly non-trivial to converge the EXX calculations in that case.

\begin{figure}[tb]
   \includegraphics[width=4.2 cm]{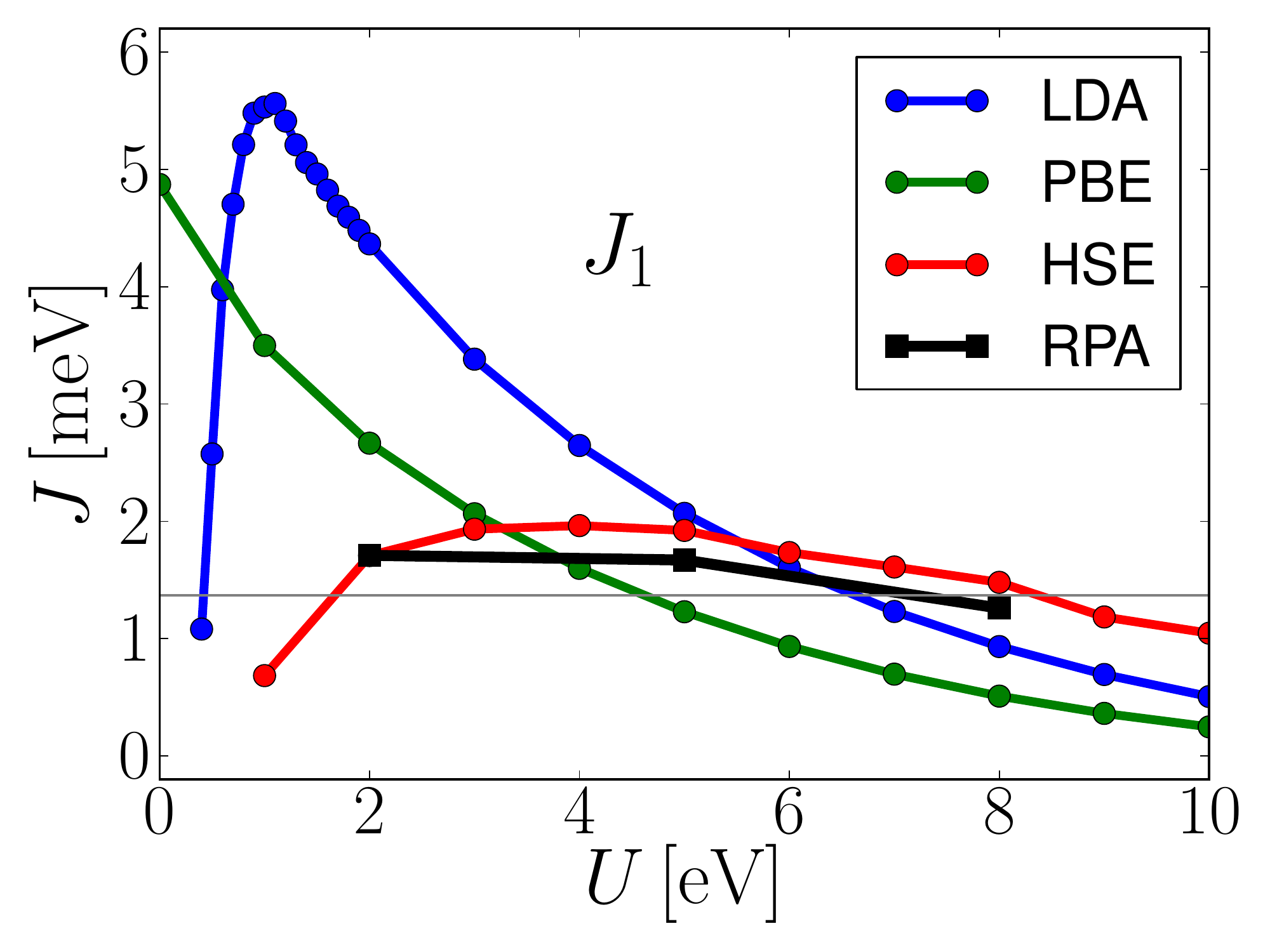}
   \includegraphics[width=4.2 cm]{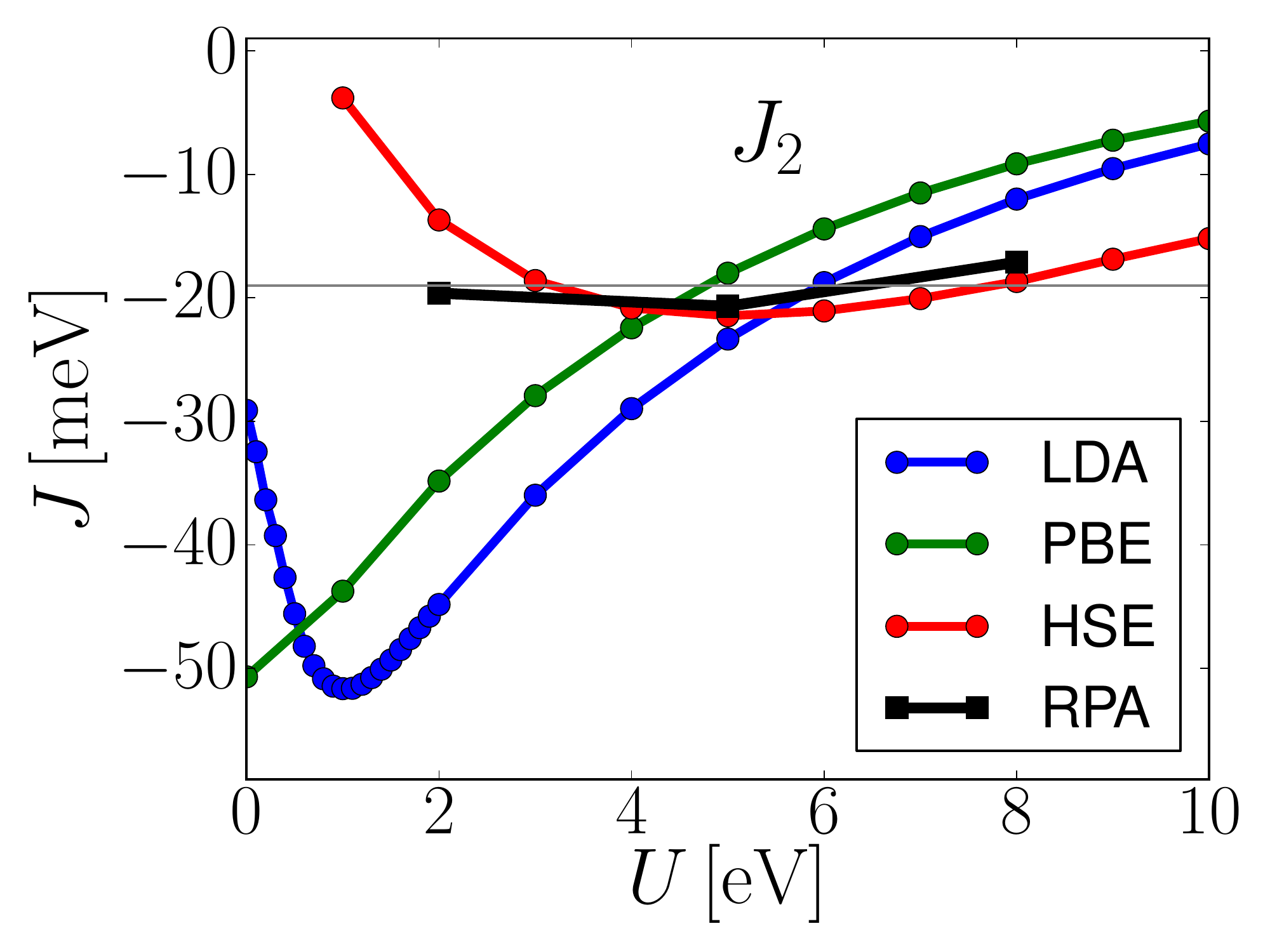}
\caption{(Color online) $J_1$ and $J_2$ for NiO calculated as a function of U using LDA+U, PBE+U, HSE06@PBE+U and RPA@PBE+U. The experimental values\cite{Hutchings1972} are indicated with grey horizontal lines.}
\label{fig:NiO}
\end{figure}

\begin{figure}[tb]
   \includegraphics[width=4.2 cm]{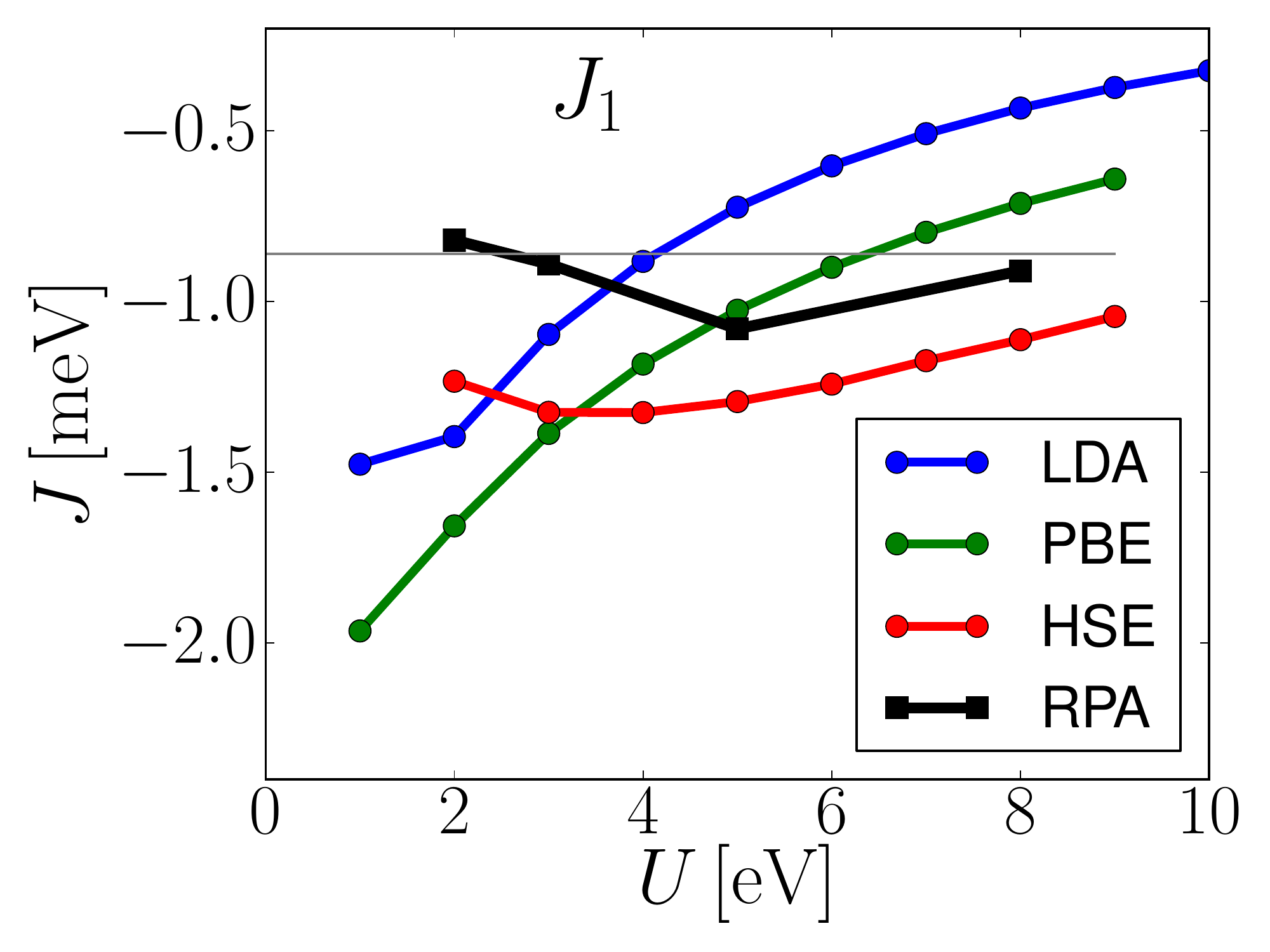}
   \includegraphics[width=4.2 cm]{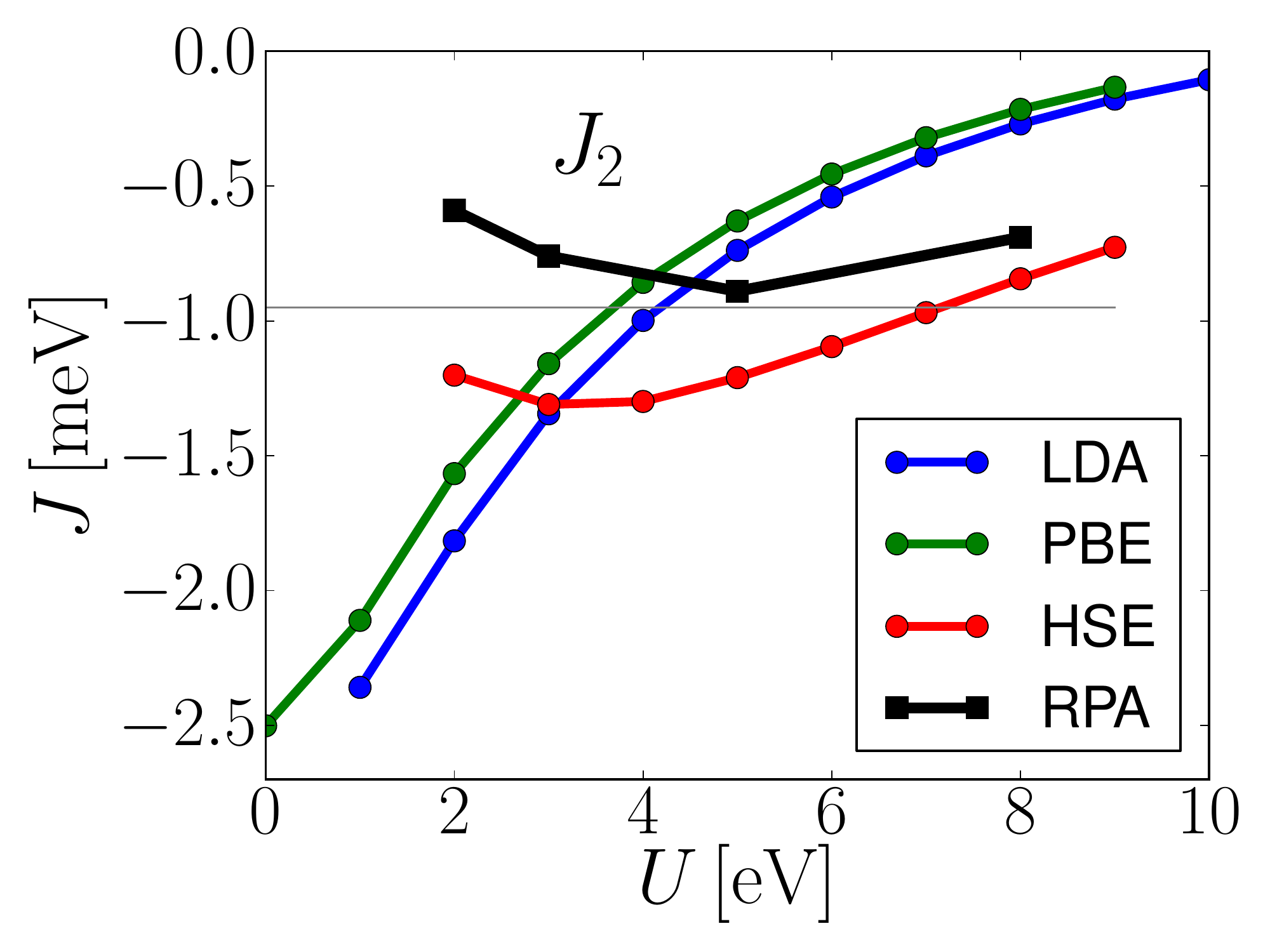}
\caption{(Color online) $J_1$ and $J_2$ for MnO calculated as a function of U using LDA+U, PBE+U, HSE06@PBE+U and RPA@PBE+U. The experimental values\cite{Lines1965} are indicated with grey horizontal lines.}
\label{fig:MnO}
\end{figure}

Although non-selfconsistent EXX completely fails in the description of the exchange parameters $J_{ij}$, standard Hubbard-corrected local and semi-local functions such as LDA + U and PBE + U, are able to give a good account of the interactions if U is chosen correctly. This is shown in Fig. \ref{fig:NiO} and \ref{fig:MnO} for the case of NiO and MnO respectively. In both cases RPA provides good agreement with experimental values, but so do LDA+U and PBE+U if one chooses U in the vicinity of 5 eV. However, both of these functionals show a strong dependence on U and all coupling parameters tend to vanish as U is increased. Nevertheless, LDA+U and PBE+U calculations are one to two orders of magnitude faster than RPA calculations and may comprise a practical choice if one is able to judiciously choose a suitable value of U. Finally, the HSE06 functional gives a very similar trend as RPA, with a shallow extremum when U is close to 5 eV. 

\subsection{Na$_3$Cu$_2$SbO$_6$}
The Na/Cu antimonate Na$_3$Cu$_2$SbO$_6$\cite{Smirnova2005} has attracted a sizable amount of interest due to evidence of a quasi-1D nature of the magnetic structure.\cite{Miura2007} Briefly, the Cu$^{2+}$ ions are arranged in a distorted two-dimensional honeycomb lattice and thus comprise a spin-1/2 system. The distortion results in the Cu atoms forming dimers with intra-dimer coupling $J_1$. In addition, the dimers interact by an inter-dimer coupling $J_2$ and the system can be regarded as a quasi-1D chain of dimers interacting through $J_2$. The sign of the inter-dimer coupling $J_1$ are of fundamental importance in order to understand the spin dynamics in the system,\cite{Schmitt2014} but there has been some controversy regarding the sign of $J_1$\cite{Derakhshan2007,Schmitt2014} as both antiferromagnetic and ferromagnetic values of $J_1$ are consistent with experimental data. In Ref. \onlinecite{Schmitt2014} the values $J_1=17.8$ meV and $J_2=-15.0$ meV were obtained by fitting quantum Monte Carlo simulations to the measured susceptibility and we will take these values as a reference here. 

\begin{figure}[tb]
   \includegraphics[width=4.2 cm]{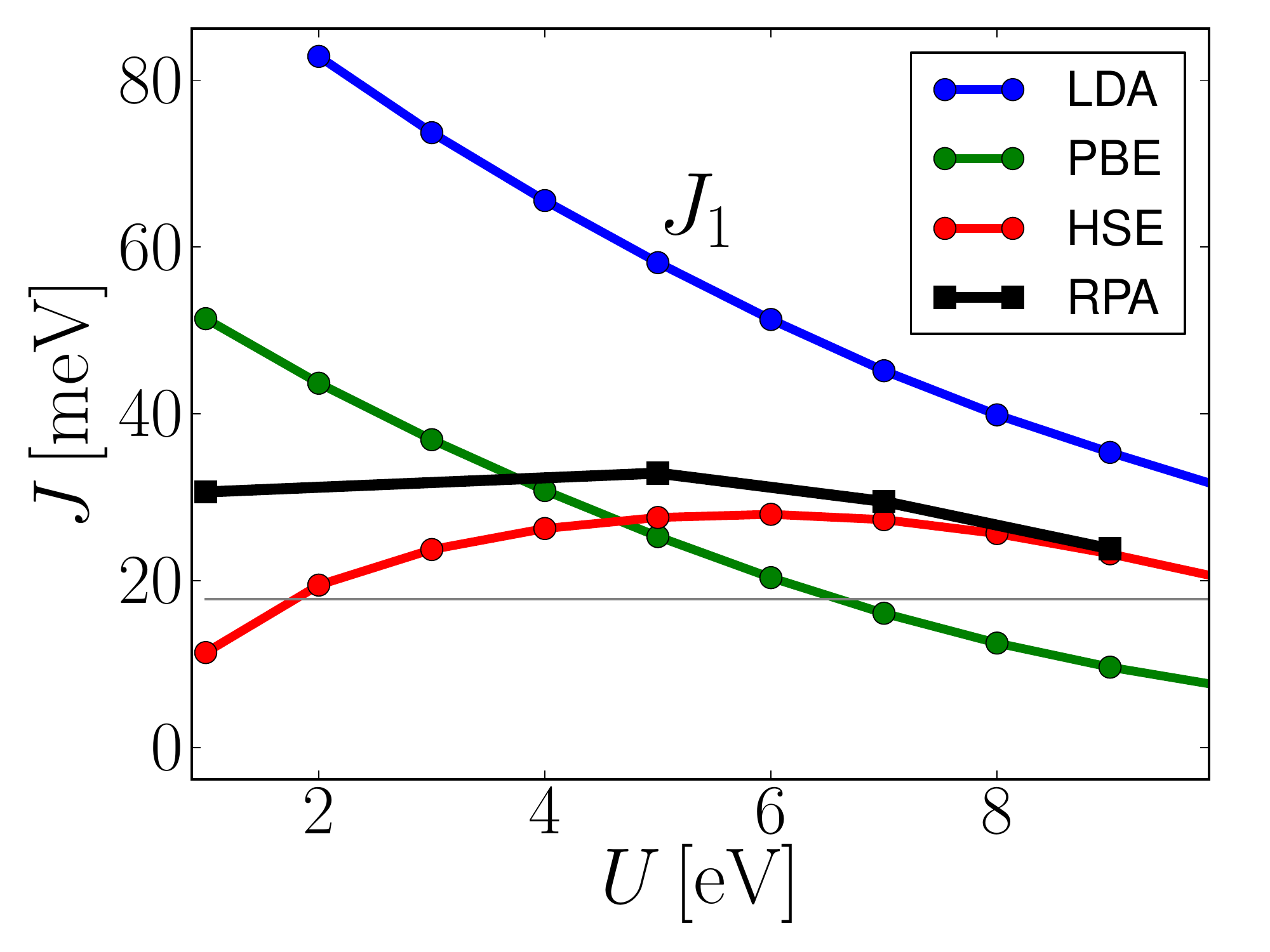}
   \includegraphics[width=4.2 cm]{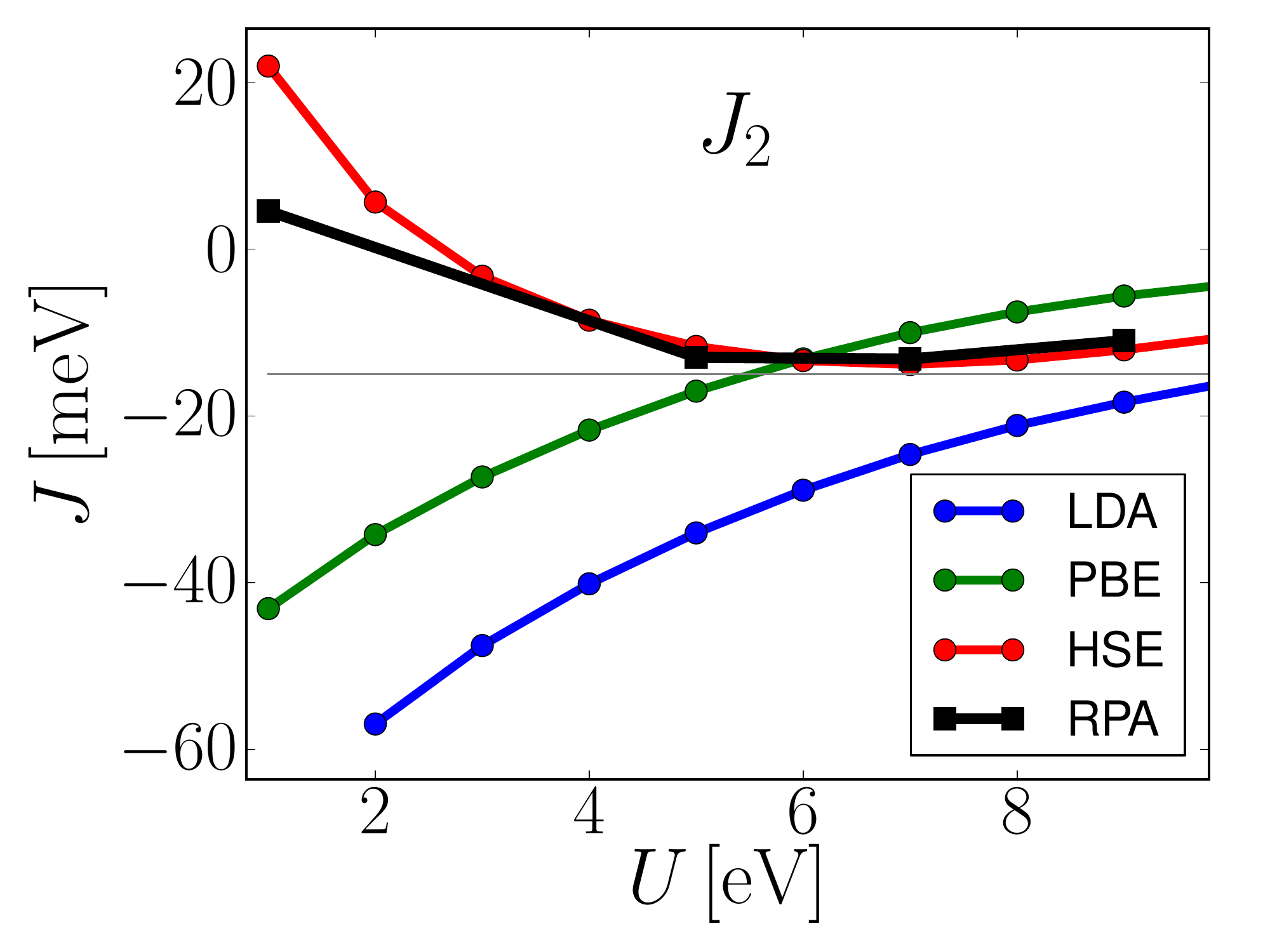}
\caption{(Color online) $J_1$ and $J_2$ for Na$_3$Cu$_2$SbO$_6$ calculated as a function of U using LDA+U, PBE+U, HSE06@PBE+U and RPA@PBE+U. The experimental values\cite{Schmitt2014} are indicated with grey horizontal lines.}
\label{fig:Na}
\end{figure}

In order to map out the energies of different spin configurations we used a unit cell containing two units of Na$_3$Cu$_2$SbO$_6$ and a gamma-centered k-point grid of $4\times4\times2$. In Fig. \ref{fig:Na} we show the coupling constants calculated with LDA+U,PBE+U,HSE06 and RPA and we see that all calculations except HSE06 at low values of U predict a ferromagnetic intra-dimer coupling. Again RPA is seen to be much less sensitive to the value of U and generally agrees well with quantum Monte Carlo simulations fitted to experiment.\cite{Schmitt2014}

\subsection{Sr$_2$CuO$_3$}
Superconductivity in the strongly correlated cuprates are suspected to be closely related to magnetic interactions and it is thus of vital importance that reliable values of the exchange coupling constants can be obtained in these materials. Here we will consider the case of Sr$_2$CuO$_3$, which exhibit strong magnetic interactions. In this material the Cu planes form 1D chains of Cu atoms connected by oxygen bridges that facilitate superexchange interaction and the magnetic structure is thus dominated by a single $J$ parameter. The experimentally determined values of $J$ range from -0.241 eV from inelastic neutron scattering\cite{Walters2009} to -0.146 eV based on susceptibility measurements.\cite{Ami1995} In Ref. \onlinecite{Foyevtsova2014} a value of -0.159 eV was obtained from extensive diffusion Monte Carlo simulations. 

\begin{figure}[tb]
   \includegraphics[width=7.0 cm]{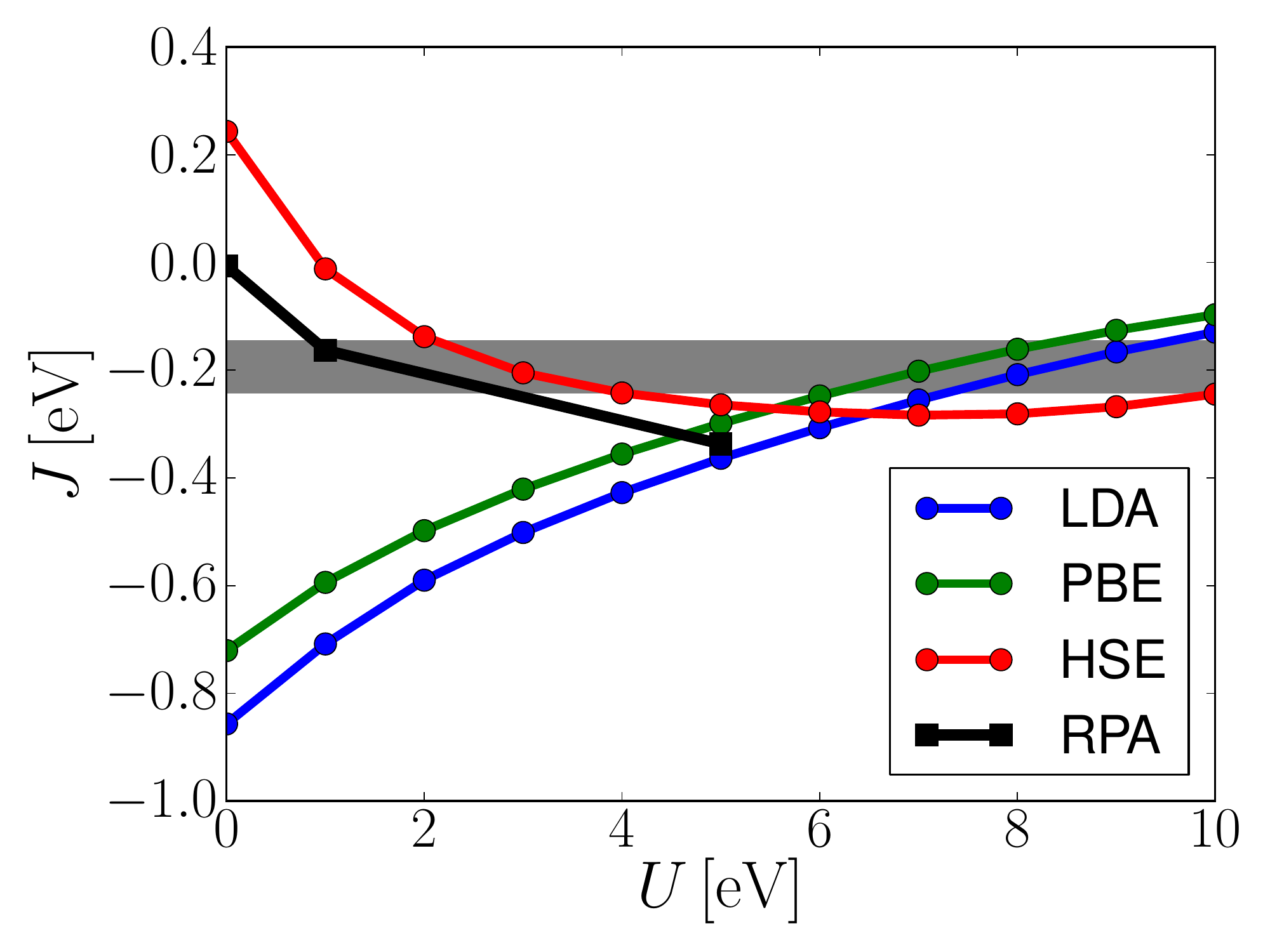}
\caption{(Color online) Nearest neighbor coupling $J$ for Sr$_2$CuO$_3$ calculated as a function of U using LDA+U, PBE+U, HSE06@PBE+U and RPA@PBE+U. The experimental span of values is indicated by the grey area. (Lower bound: inelastic Neutron scattering.\cite{Walters2009} Upper bound: susceptibility measurements.\cite{Ami1995})}
\label{fig:Sr}
\end{figure}

In order to map out the energies of the two different spin configurations needed to calculate $J$ we used a unit cell containing four units of Sr$_2$CuO$_3$ and a gamma-centered k-point grid of $6\times6\times2$. We present the results obtained from LDA+U PBE+U, HSE@PBE+U, and RPA@PBE+U in Fig. \ref{fig:Sr}. Again we see the trend that LDA+U and PBE+U provides a good approximation for U in the range 5-10 eV, but severely overestimates the magnitude of $J$ for small values of U. RPA shows less sensitivity to to U, but seems to overestimate $J$ for intermediate values of U. HSE06 follows the same trend as RPA, but gives the wrong sign of $J$ for U$<1$.

\subsection{Sr$_2$CuTeO$_6$}
Another magnetic material that has recently be scrutinized theoretically\cite{Babkevich2016} is the double perovskite Sr$_2$CuTeO$_6$. Here the magnetic Cu$^{2+}$ ions comprise 2D square planes where the superexchange interactions are mediated by intermediate TeO$_6$ octahedra, which make the interactions rather weak compared to oxygen mediated superexchange. In addition to nearest neighbor antiferromagnetic interactions $J_1$ there is also next-nearest antiferromagnetic interactions $J_2$ that introduces a frustration in the 2D magnetic lattice and is thus a prime example of a frustrated 2D magnet on the rectangular lattice. Due to the strongly correlated nature of the material and the very weak interactions it has proven highly challenging to faithfully reproduce coupling constants determined from inelastic neutron scattering. However, in Ref. \onlinecite{Babkevich2016} it was demonstrated that agreement with experiments could obtained with sophisticated quantum chemistry methods. 

\begin{figure}[tb]
   \includegraphics[width=4.2 cm]{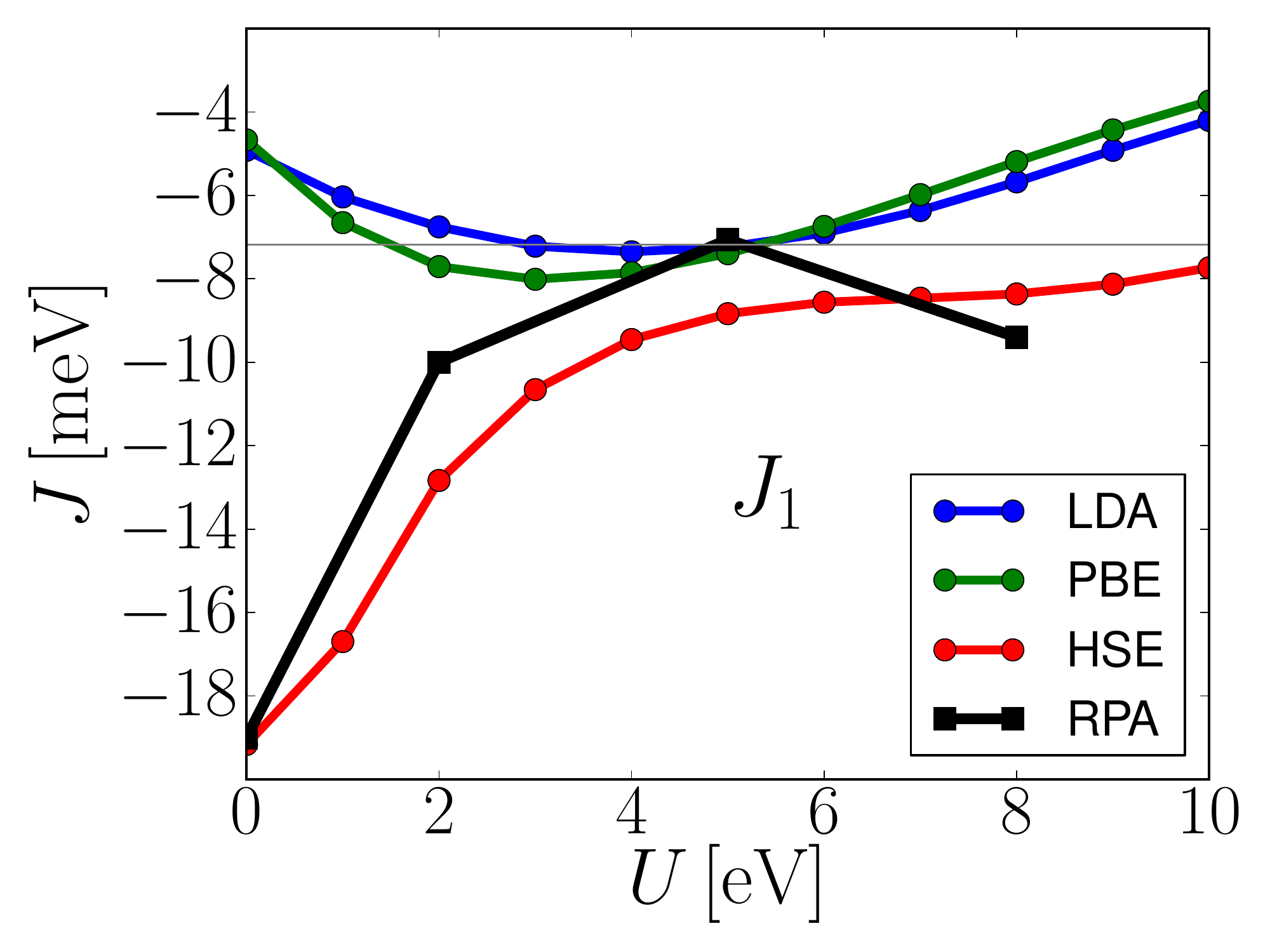}
   \includegraphics[width=4.2 cm]{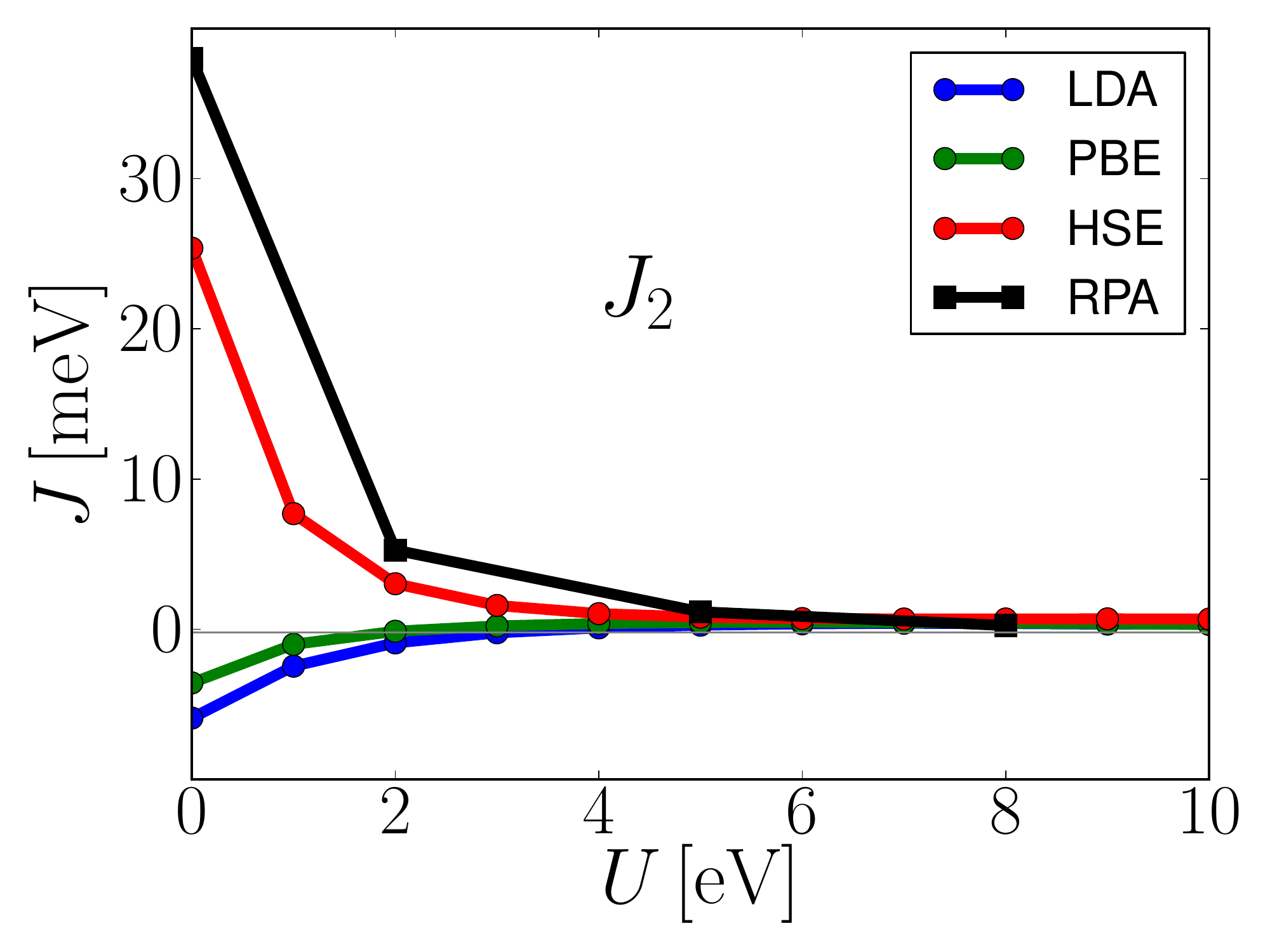}
\caption{(Color online) $J_1$ and $J_2$ for Sr$_2$CuTeO$_6$ calculated as a function of U using LDA+U, PBE+U, HSE06@PBE+U and RPA@PBE+U. The experimental values\cite{Babkevich2016} are indicated with grey horizontal lines.}
\label{fig:SrTe}
\end{figure}

We used a unit cell containing four units of Sr$_2$CuTeO$_6$ and a total of 16 k-points to construct the three spin configurations needed to evaluate $J_1$ and $J_2$. In Fig. \ref{fig:SrTe} we present the results and it is evident that RPA and HSE have severe problems in capturing the correct interactions for small values of U. In particular, RPA and HSE06 give the wrong sign of $J_2$ , which is experimentally determined to be -0.21 meV. LDA+U and PBE+U on the other hand gives the correct sign of $J_2$ for small values of U, but becomes positive for U $>$ 3 eV.

\subsection{Monolayer of CrI$_3$}
\begin{figure}[tb]
   \includegraphics[width=7.0 cm]{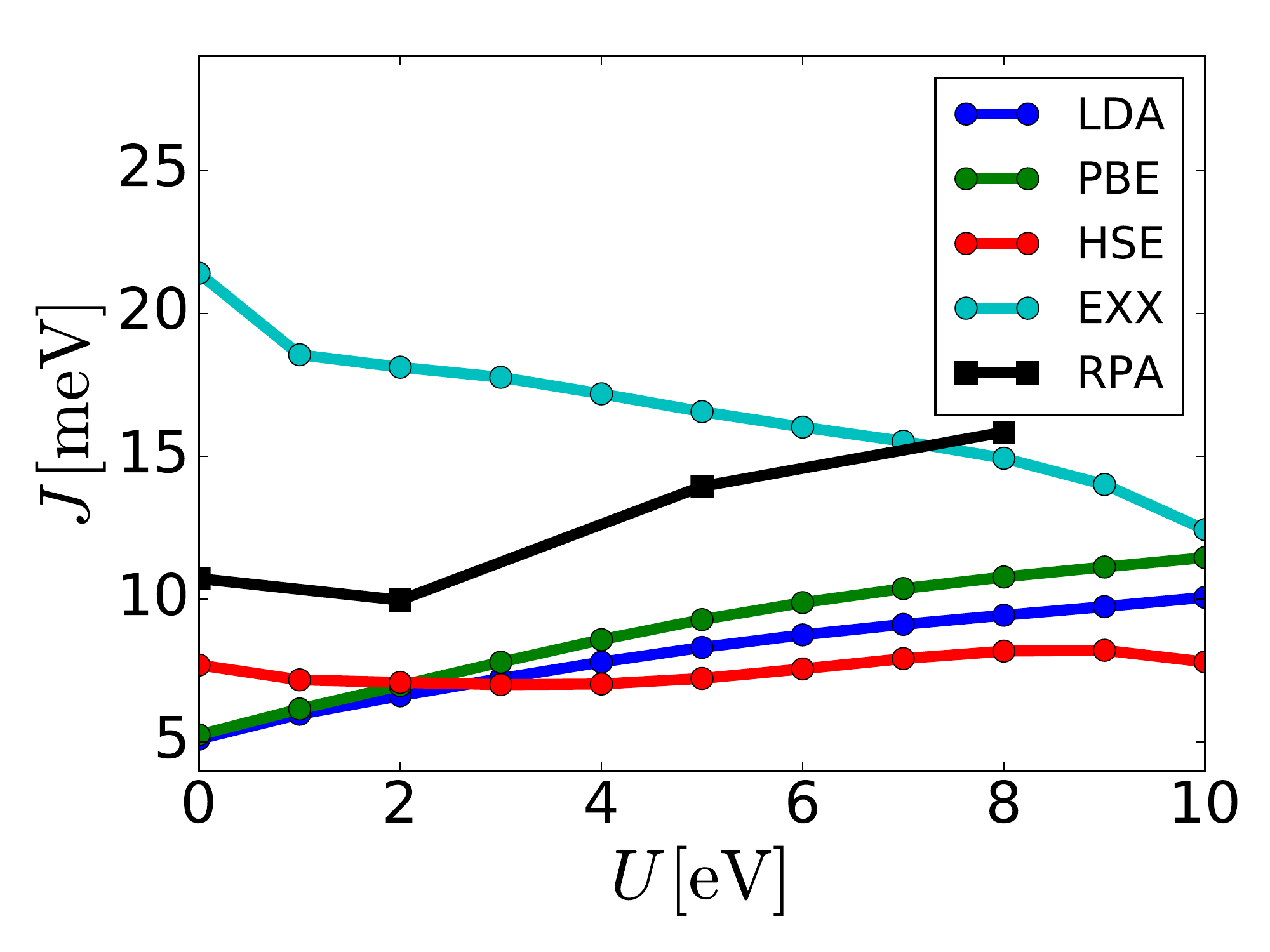}
\caption{(Color online) Nearest neighbor coupling $J$ for a monolayer of CrI$_3$ calculated as a function of U using LDA+U, PBE+U, EXX@PBE+U, HSE06@PBE+U and RPA@PBE+U.}
\label{fig:Cr}
\end{figure}

The simple Heisenberg model Eq. \eqref{eq:Heisenberg} cannot accommodate magnetic order in a purely two-dimensional system according to the Mermin-Wagner theorem. However, if one includes the effects of either magnetic anisotropy or a quasi-2D material it becomes possible to bypass the Mermin-Wagner theorem as was recently demonstrated experimentally for CrI$_3$\cite{Huang2017} and a bilayer of Cr$_2$Ge$_2$Te$_6$\cite{Gong2017a} respectively. The possibility of magnetic order in two-dimensional systems constitutes an intriguing route to construct nano-scale spintronics devices and accurate determination of the magnetic interactions and magnetic anisotropy in such systems are crucial prerequisites for a faithful prediction of the magnetic properties. Here we consider a monolayer of CrI$_3$ as an example of a pure 2D magnetic system. The magnetic Cr ions form a honeycomb lattice such that each ion has three nearest neighbors and we will neglect all but the nearest neighbor interactions and consider only a single magnetic interaction $J$. In Fig. \ref{fig:Cr} we show the calculated value $J$ as a function of $U$. Once again, LDA+U and PBE+U give rise to monotonously increasing values of $J$ when U is increased. HSE06@PBE+U on the other hand is seen to be largely independent of the value of U, but yield values of $J$ similar to LDA and PBE, whereas RPA@PBE+U yield much larger values of $J$. In particular, it has been shown that the electronic properties of Cr-based halides \cite{Chen2017a} are well reproduced with U=0 and at this point RPA yields $J=$ 10.5 meV, whereas LDA and PBE give $J=$ 5 meV. Thus the magnetic interaction differ by a factor of two when comparing LDA and PBE with RPA, which is bound to have a crucial influence on the predicted values of the Curie temperature for this material. 

The fact that $J$ increases with larger values of $U$ indicate that the superexchange plays an important role although this material seems to be dominated by direct exchange at first sight. Larger values of $U$ tend to localize the orbitals carrying the magnetic moments and thus decrease the hybridization that gives rise to direct exchange. This is also confirmed by EXX@PBE+U calculations in Fig. \ref{fig:Cr}.  There is presently no reliable experimental estimates of $J$ for this material so with the data at hand it is not possible to determine whether the semi-local functionals or RPA provides to most accurate prediction. Nevertheless, the calculations demonstrate that one should be careful in trusting the accuracy of semi-local functionals for magnetic interactions in this case. Based on the failure of LDA and PBE at U=0 for magnetic interactions of the other materials studied in this work, it is likely that RPA provides a better estimate of $J$ for this material and perhaps two-dimensional materials in general.

Regardless of the values of $J$ magnetic order can only exist in a purely two-dimensional material if magnetic anisotropy is present. We have calculated the anisotropy by including the spin-orbit coupling non-selfconsistently\cite{Olsen2016a} and obtain an energy difference between magnetic moment in plane and out-of plane of 0.9 meV per Cr atom. The easy axis is out of plane, which is also what is found experimentally.\cite{Huang2017}

\section{Conclusion}\label{sec:discussion}
We have performed first principles calculations of magnetic interactions in the magnetic materials NiO, MnO, Na$_3$Cu$_2$SbO$_6$, Sr$_2$CuO$_3$, Sr$_2$CuTeO$_6$, and a monolayer of CrI$_3$, within LDA+U, PBE+U and HSE06@PBE+U and RPA@PBE+U. The RPA can be decomposed into an EXX part and a correlation part and as expected the EXX part fails dramatically due to the presence of the superexchange mechanism in these materials. Moreover, the results are highly dependent of the value of U used in the calculations. In contrast, RPA is able to accurately correct EXX results for a wide range of the U parameters. For the classic Mott insulators RPA provides very good agreement with experiments, whereas LDA+U and PBE+U only provides accurate values when choosing U$\sim$5 eV. A similar picture arises for the materials Na$_3$Cu$_2$SbO$_6$ and Sr$_2$CuO$_3$ where RPA gives a much better estimate of the magnetic interactions for small values of U than LDA+U and PBE+U, whereas PBE+U gives very good results for U$\sim$5 eV. For the frustrated magnet Sr$_2$CuTeO$_6$, RPA seems to perform rather poorly for low values of U whereas it coincides with LDA+U and PBE+U for U$\sim$5 eV - close to the experimental values. Finally, RPA predicts much stronger magnetic interactions in the recently discovered two-dimensional magnet CrI$_3$ than either LDA+U, PBE+U or HSE06.

To conclude, EXX has been shown to fail dramatically whenever magnetic interactions are not describe by idealized direct exchange. This has been the case for all the materials studied in the present work - even the monolayer of CrI$_3$, which look like a simple exchange-mediated ferromagnet at first sight. RPA incorporates the correlations necessary to capture the superexchange mechanism and effectively brings the predicted values of ferromagnetic and antiferromagnetic magnetic interactions close to experimental values.  However, in many cases, similar accuracy can be obtained with PBE+U, which is much less demanding from a computational point of view. Applying PBE+U, however, requires prior knowledge of suitable values of U for particular materials. When suitable values of U is not known or fixed by other requirements such as values for the band gap and magnetic anisotropy, RPA could constitute a valuable parameter-free alternative.

\section{Acknowledgments}
The Center for Nanostructured Graphene (CNG) is sponsored by the Danish National Research Foundation, Project DNRF58.



%

\end{document}